\documentclass{article}
  
\usepackage{graphicx}
\usepackage{latexsym} 
\usepackage{amsmath}
\def\bgeq{\begin{equation}} \def\edeq{\end{equation}}
\def\bgeqy{\begin{eqnarray}} \def\edeqy{\end{eqnarray}}

\def\sn{{\smallskip\par\noindent}}
\def\refeq#1{{Eq.~(\ref{#1})}}
\def\reffg#1{{Fig.~\ref{#1}}} \def\reffgs#1{{Figs.~\ref{#1}}}
  
\begin{document}

%\classification{}
\title{Data Processing Approach for Localizing Bio-magnetic Sources in the
Brain}

%\author{Hung-I Pai$^{1\dagger}$, Chih-Yuan Tseng$^{1}$ and 
%H.C. Lee$^{1,2,3}$ \\
%EndAName
%\small{$^{1}$Computational Biology Laboratory}\\
%\small{$^{1}$Department of Physics, National Central University, 
%Zhongli, Taiwan 32001}\\
%\small{$^{2}$Graduate Institute of Systems Biology and Bioinformatics}\\
%\small{National Central University, Zhongli, Taiwan 32001}\\
%\small{$^{3}$National Center for Theoretical Sciences, Hsinchu, Taiwan 30043}\\
%}{}

\author{Hung-I Pai$^{1}$\thanks{He is currently working in Industrial Technology Research Institute, Chu Tung, Hsin Chu, Taiwan 310.}, Chih-Yuan Tseng$^{1}$\thanks{He is currently at Department of Oncology, University of Alberta, Edmonton AB T6G 1Z2 Canada.}
and H.C. Lee$^{1,2,3}$ \\
\small{$^{1}$Computational Biology Laboratory}\\
\small{$^{1}$Department of Physics, National Central University, Zhongli, Taiwan 32001}\\
\small{$^{2}$Graduate Institute of Systems Biology and Bioinformatics}\\
\small{National Central University, Zhongli, Taiwan 32001}\\
\small{$^{3}$National Center for Theoretical Sciences, Hsinchu, Taiwan 30043}\\}
\date{}
\maketitle

\begin{abstract}
Magnetoencephalography (MEG) provides dynamic spatial-temporal insight of 
neural activities in the cortex. Because the number of possible sources is
far greater than the number of MEG detectors, the proposition to localize
sources directly from MEG data is notoriously ill-posed. Here we develop an approach 
based on data processing procedures including 
clustering, forward and backward filtering, and the method of maximum 
entropy.  We show that taking as a starting point the assumption that the sources 
lie in the general area of the auditory cortex (an area of about 40 mm by 15 mm), 
our approach is capable of achieving reasonable success in pinpointing active sources 
concentrated in an area of a few mm's across, while limiting the spatial distribution and 
number of false positives.   
\end{abstract}


%%%%%%%%%%%%%%%%%%%%%%%%%%%%%%%%%%%%%%%%%%%%
%% MAINMATTER
%%%%%%%%%%%%%%%%%%%%%%%%%%%%%%%%%%%%%%%%%%%%

\section{Introduction}

Magnetoencephalography (MEG) records magnetic fields generated from
neurons when the brain is performing a specific function. Neural
activities thus can be noninvasively studied through analyzing the MEG
data. Since the number of neurons (unknowns) are far larger than the
number of MEG sensors (knowns) outside the brain, the problem 
of identifying activated neurons from the magnetic data is ill posed. 
The problem becomes even more severe when noise is present. 

%%%%%%%%%%%% {f:testMRIregion} %%%%%%%%%%%%%%
\begin{figure}[htb]
\center
\includegraphics[height=1.4in]{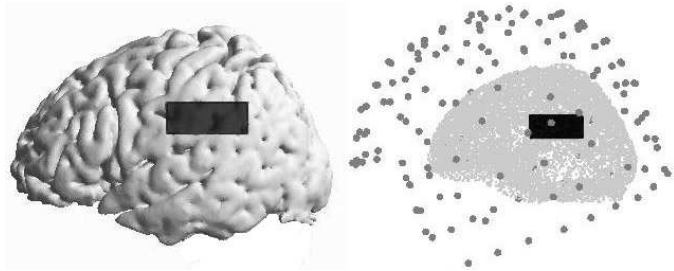}%{fig1v3}
\caption{Left, side view of the human cortex; the frontal lobe points to the left.  
The region of interest, the auditory cortex, is marked by  the dark rectangular.  
Right, schematic setup of an MEG experiment.  The occipital lobe (right side in graph) 
is closer to the sensors because the person being tested is lying face-up.}
  \label{f:testMRIregion}
\vspace{-0.5cm}
\end{figure}
%%%%%%%%%%%%%%%%%%%%%%%%%%%%%
A first and essential step in surpassing the obstacle of ill-posedness is to 
rely on prior knowledge of the general area of 
active current sources producing the MEG data. 
Often, this prior knowledge is provided by functional magnetic resonance imaging (fMRI) 
experiments.  The two kinds of experiments complement each other, 
MEG has high temporal resolution (about $10^{-3}$s) but poor spatial resolution, 
whereas fMRI has high spatial resolution \cite{Moonen99} but poor temporal resolution (about 1 s).
Consider auditory neural activity.   FMRI shows that such activity is concentrate 
in the auditory cortex, two 40 mm by 15 mm areas respectively located on the two 
sides of the cortex (\reffg{f:testMRIregion}L).  
Because of poor temporal resolution, fMRI cannot resolve the 
rapid successive firing of the (groups of) neurons, instead it shows large areas of 
the auditory cortex lighting up.  Under MEG, the same auditory activity will be detected 
by high time-resolution sensors (\reffg{f:testMRIregion}R)
which however collectively cover an area (when projected 
down to the cortex) much greater than the auditory cortex.    

Although many methods have been proposed to attack the ill-posed
problem, including dipole fitting, minimum-norm-least-square (MNLS)
\cite{Method:MNLS}, and the maximum entropy method (ME) \cite{ME:ME},
improvements have been limited.  One reason for this may be
that these methods do not properly include anatomic constrains. In
this work, we propose a novel approach to analyze noisy MEG data 
based on ME that pays special attention to obtaining better {\it priors} as input to the 
ME procedure by employing clustering and forward and backward filtering processes 
that take anatomic constrains into consideration.   
We demonstrate the feasibility of our approach by testing it on several simple cases. 

\section{The ill-posed inverse problem}
Given the set of current sources (dipole strengths) \{$r_i\vert i=1,2,\cdots,N_r$\} at sites 
\{$z_i\vert i=1,2,\cdots,N_r$\},   
the magnetic field strength $m(\mathbf{x})$ 
measured by the sensor at spatial position {\bf x} is given by, 
\begin{equation}
m(\mathbf{x})=\sum_{i=1}^{N_r}A(\mathbf{x})_i r_i + n(\mathbf{x})
\equiv {\hat A}(\mathbf{x})\cdot {\hat r}+ n(\mathbf{x})
\label{Eq:M=Ar}
\end{equation}%
where the function $A(\mathbf{x})_i$, which is a function of $z_i$,  
is derived from the Biot-Savart law in vacuum, 
$n(\mathbf{x})$ is a noise term, and a hatted symbol denotes an 
$N_r$-component vector in source space.   We consider the case where there are 
 $N_m$ sensors at locations {\bf x}$_\alpha$, $\alpha$=1,2,$\cdots,N_m$. 
To simplify notation, we write $\mathbf{m}$ as an $N_m$-component vector 
whose $\alpha^{th}$ component is $m^\alpha$=$m(\mathbf{x}_\alpha)$, similarly for 
$\mathbf{\hat A}$  and $\mathbf{n}$.   Then Eq. (\ref{Eq:M=Ar}) simplifies to 
\bgeq
\mathbf{m} = \mathbf{\hat A}\cdot {\hat r} + \mathbf{n}
\label{e:inverse}
\edeq
In practice, 
magnetic field strengths are measured at the $N_m$ sensors and the inverse 
problem is to obtain the set of $N_r$ dipole strengths $r_i$ with $N_r>>N_m$.  
The presence of noise 
raises the level of difficulty of the inverse problem.  
\section{Methods}

\sn \textbf{Human Cortex and MEG Sensors}. 
In typical quantitative brain studies, the approximately 10$^{10}$ neurons in the cortex 
are simulated by about 2.4$\times$10$^{5}$ current dipoles whose  directions are set parallel to 
the normal of the cortex surface \cite{MRI:database,MRI:orientation}. 
For this study we focus on the auditory cortex, the area marked 
by the 40 mm$\times$ 15 mm rectangular shown in \reffg{f:testMRIregion}L 
that contains 2188 current dipoles.  

In typical MEG experiments the human head is surrounded by a hemispheric tiling of magnetic field 
sensors called superconducting quantum interference devices (SQUIDS). In the experiment 
we consider,  there are 157 sensors, each composed of a pair of co-axial 
coils of 15.5 mm diameter 50 mm apart, with the lower coil 
about 50 mm above the scalp \cite{Kado99}.
The centers of the sensors are represented by the gray dots in \reffg{f:testMRIregion}R.  
Details of geometry and of the A-matrixes in \refeq{Eq:M=Ar} are given in \cite{Details}. 

\begin{figure}[hbp]
\center
\includegraphics[height=1.8in]{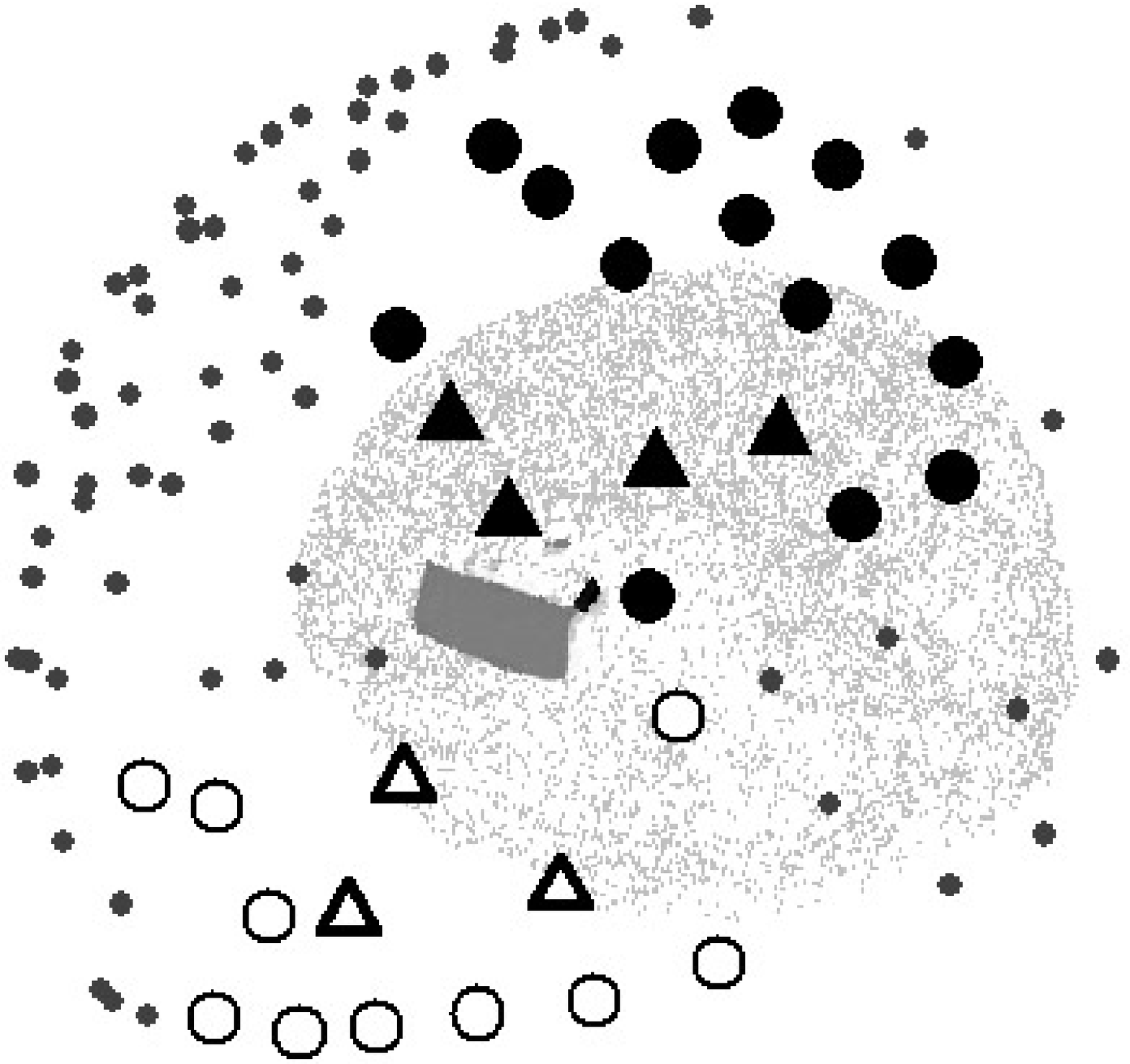}%{fig2v5}\hspace{20pt} 
\includegraphics[width=2.2in]{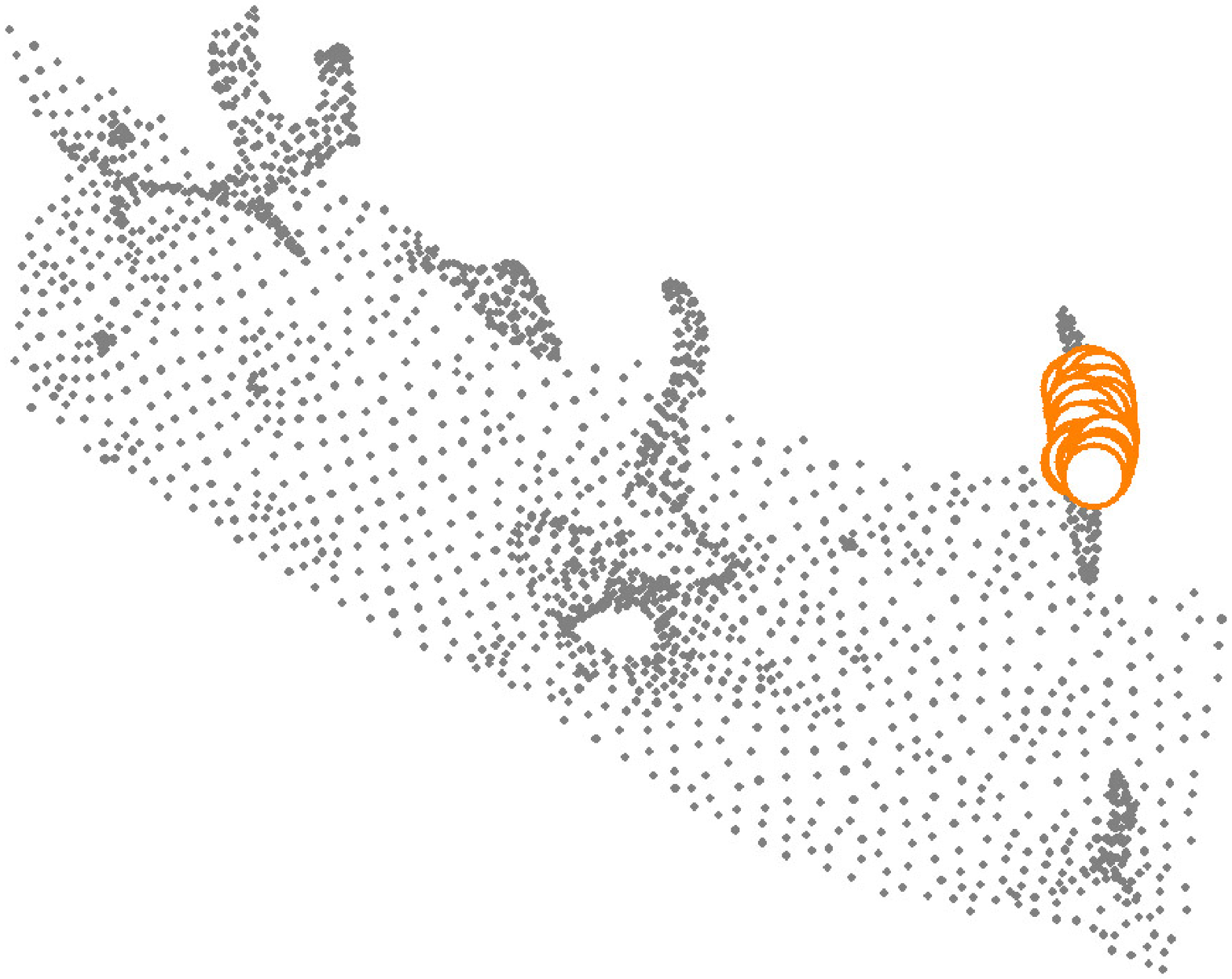}
  \caption{Left, the circular and triangular symbols 
  are the positions of the 31 sensors with detectable 
  signals, including the 7 (triangular) with signals above threshold.  
  Sensors with magnetic flux going into (out) the page are solid (hollow). 
 Right, detail of auditory cortex. The dark (orange in color) circles in the top-right corner 
 indicate the general area of active sources used to generate artificial MEG data.}
\label{f:M_sensors}
\end{figure}

\sn \textbf{Artificial MEG data and Noise}. 
We use artificial MEG data generated by the forward equation, 
Eq. (\ref{Eq:M=Ar}), from sets of current dipoles (to be specified below) in a small 
area (black circles in \reffgs{f:M_sensors}R) within the auditory cortex 
(the gray "rectangular" region in \reffgs{f:M_sensors}L, enlargement shown  
in \reffgs{f:M_sensors}R).  
%%%%%%%%%%% {f:M_sensors} %%%%%%%%%%%
%%%%%%%%%%%%%%%%%%%%%%
A site-independent white Gaussian noise is linearly superimposed on the MEG data.
The signal to noise ratio (SNR) is defined as
\bgeq
SNR=-10log_{10}||\mathbf{n}_{max}||^2/||\mathbf{m}_{max}||^2
\edeq
where $\mathbf{m}_{max}$ is the amplitude at the sensor receiving the strongest 
noiseless MEG signal, and $\mathbf{n}_{max}$ is amplitude of the strongest simulated noise. 
In this study we have $\mathbf{m}_{max}$=7.4 fT (fT= femto-Tesla) and 
$\mathbf{n}_{max}$=0.05$\mathbf{m}_{max}$=0.37 fT, so that $SNR$=26 on each 
individual run.  
The artificial MEG data is generated by giving a current of 10 nA (nano-ampere) to each of 
the sources in a source set (see below), running the forward equation with noise 
10 times and taking the averaged strengths at the sensors.   The averaging has the effect 
of reducing the effective $\mathbf{n}_{max}$ by a factor of $\sqrt{10}$, and yielding an 
enhanced effective signal to noise ratio of $SNR'$=36. 

Using a threshold of $T_S$=14$\mathbf{n}_{max}$=5.1 fT we select a subset $M_S$ 
of 7 "strong signal" sensors.   This implies a minimum value of $SNR'$=32.9 
(with averaging) on each sensor in the set.  
Given the (assumed) normal distribution of noise intensity, this selection 
implies that at 99.99$\%$ confidence level the signals considered are not noise. 
Similarly we use a threshold of $T_C$=6$\mathbf{n}_{max}$=2.2 fT to select a 
subset $M_C$ of 31 "clear signal" sensors, with minimum values of $SNR'$=25.6 
on each sensor in the set.  
In actual computations below, we reduce the sensor space to one that include only 
those in the set $M_C$.  In practice, the reduction replaces $\mathbf{m}$, $\mathbf{A}$, 
and $\mathbf{n}$ by $\mathbf{m'}$, $\mathbf{A'}$, and $\mathbf{n'}$, respectively.  

\sn \textbf{Receiver Operating Characteristics Analysis}. 
We evaluate the goodness of our results using  receiver operating characteristics (ROC) 
analysis \cite{ROC},  in which the result is presented in the form of a plot of 
the true positive rate ($S_n$, sensitivity) versus the  false positive rate (or 1-$S_p$, where 
$S_p$ is the specificity).   Let $R$ be the total solution space of current sources, 
$T$ the true solutions, or actual active sources, and $P$ the positives, or the predicted 
active sources.  Then $F$=$R$-$T$ is the false solutions, 
$TP$=$(P\bigcap T)$ the true positives, $FP$=($P\bigcap T$)-$T$ the false positives, 
and $FN$=$R$-($P\bigcap T$) the false negatives.    By definition 
$S_{n}$=$TP/T$=$(P\bigcap T)/T$ and 1-$S_p$=$FP/F$=$((P\bigcup T)-T)/(R-T)$. 
Intuitively, a good solution is one such that maximizes $S_{n}$ while minimizing 1-$S_p$. 
In a null theory, the positives will fall randomly into $R$, hence $TP/T$=1-$S_p$, 
or $S_{n}$=1-$S_p$.  
Therefore, the merit of model producing a piece of data, ($S_{n}$, 1-$S_p$), showing on 
an ROC plot is measured by the difference between $S_{n}$ and 1-$S_p$.  In general, 
when a model is used to generate a curve in an ROC plot, the "area under the curve" (AUC), or the 
area between the curve and the $S_{n}$=1-$S_p$ line, is a measure of merit of the 
model \cite{ROC2}.    

\begin{figure}%[hbp]
\center
  \includegraphics[height=1.8in]{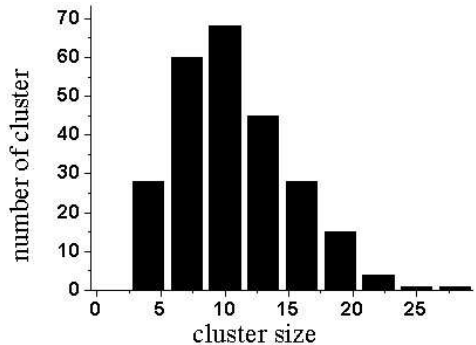}%{clusternumberv1}
  \caption{Distribution of cluster size in the clustering of 2188 sources into 250 clusters.}
  \label{f:clustersize_dist}
\end{figure}

\sn \textbf{Clustering and Sorting.} 
Although implicit in the MRI head model
introduced in Fig. \ref{f:testMRIregion} is a dramatic 
reduction of the number of neurons and the complexity of
the cortex, the remaining number of effective neurons is still far greater  
than the number of detectors. We use a clustering algorithm \cite{method:cluster} to further 
decrease the number of effective sources, in which 
the sources are partitioned according to spatial proximity and 
similarity in orientation into the set of $N_{C}$ clusters  
$C=\{C_u\vert u=1,2,\cdots,N_{C}\}$, as follows. We require sources within a cluster 
to lie with a spatial radius of 5 mm and define 
($N_u=\sum_{i\in C_{u}}$ is the number of current sources in cluster $C_u$) 
\bgeq
%\langle\mathbf{A'}\rangle_u= \sum_{i\in C_{u}}\mathbf{A'}_{i}
\mathcal{A}'_u= \sum_{i\in C_{u}}\mathbf{A'}_{i}
\label{e:center-of-mass} 
\edeq
as the "strength" of the {\bf A}-matrices in cluster $C_{u}$, 
\bgeq
d_u=\sum_{i\in C_{u}}\vert N_u\mathbf{A'}_{i} - \mathcal{A}'_u \vert/N_u
\label{e:intra-distance}
\edeq
as the "radius" -- in the space of sensors -- of cluster $C_{u}$, and 
\bgeq
D_u=\sum_{C_v\in C}| \mathcal{A}'_v- \mathcal{A}'_u|/(N_C -1)
\label{e:inter-distance}
\edeq
as the average inter-cluster distance between 
$C_{u}$ and all the other clusters.  
The clustering, including $N_C$, is determined by requiring that 
\bgeq
d_u/D_u < \gamma_C,\quad \forall\ u=1,2,\cdots,N_{C},
\label{e:threshold} 
\edeq
where $\gamma_C$ is a parameter that controls the average cluster size; 
a smaller value of $\gamma_C$ implies smaller and more numerous clusters.  
In the limit $\gamma_C$$\to$0 every cluster will consist of a single source and 
$N_C$$\to$$N_r$, or 2188 in the present case. 
A clustering obtained with $\gamma_C$=1/7 was used in this work. It 
partitions the 2188 sources into $N_C$=250 clusters, whose size distribution 
is shown in \reffg{f:clustersize_dist}.    
 %%%%%%%%%%% {f:clustersize_dist} %%%%%%%%%%%
%%%%%%%%%%%%%%%%%%%%%% 

The clustering results in the replacement of original source distribution 
by a coarse-grained distribution of virtual source-clusters whose $N_r$
$\mathbf{A'}_i$-matrices are given by $N_C$ $\mathcal{A}'_u$'s. 
The clustering reduces Eq. (\ref{e:inverse}) to 
\bgeq
\mathbf{m'} = \mathcal{\hat A}'\cdot  {\hat r_C}+ \mathbf{n'} 
\label{e:inverse_cluster}
\edeq
which has the same form as \refeq{e:inverse} except that here the hatted vectors have 
only $N_C$ components and  each of $N_C$ components in $\hat r_C$ denotes the 
strength of the current dipole representing a cluster. 

It is convenient to sort the cluster set $C$ according to the field strength of the clusters.  
Since the field strength depends on the where it is measured, the sorted order 
will be sensor-dependent.  We denote the sorted set for sensor $\alpha$ by $C^{\{\alpha\}}$. 
Thus we have: 
\bgeq
C^{\{\alpha\}}=\{C_{u_\alpha}\vert u_\alpha=1,2,\cdots,N_{C}\},\quad
\vert \mathcal{A}'_{u_\alpha}\vert 
\ge \vert\mathcal{A}'_{v_\alpha}\vert
\ {\rm if}\ u_\alpha < v_\alpha, \quad \forall \alpha\in M_C.   
\label{e:order}
\edeq

\sn \noindent \textbf{Forward Filtering}.   
A key in improving the quality of the solution of an inverse problem is to reduce the 
number of false positives. 
In the MEG experiments under consideration, the plane of the sensors are generally parallel 
to the enveloping surface of the cerebral cortex. Such sensors are meant to detect signals emitted 
from current sources in sulci on the cortex, and are not sensitive to signals from sources in gyri.  
In practice, in our test cases $T$ will be composed of sulcus sources.  Therefore,  if we 
simply remove those clusters having the weakest strengths, we will reduce $FP$ at 
a higher rate than $TP$.   

Given a positive fractional number $\xi<1$, we use it to set an integer number $N_\xi <N_C$, 
and use $N_\xi$ to define the truncated sets 
\bgeq
R_\xi^{\{\alpha\}}=\{C_{u_\alpha}\vert u_\alpha=1,2,\cdots,N_\xi\},\quad
\forall\ \alpha\in M_S. 
\label{e:reduced}
\edeq  
The integer $N_\xi$ is determined by regression by demanding the union set 
\bgeq
R_\xi=\bigcup_{\alpha\in M_S} R_\xi^{\{\alpha\}} = \xi R
\label{e:R_MSM}
\edeq
to be a fraction $\xi$ of $R$.  We call this forward filtering process of reducing the pool of possible 
positives from $R$ to $R_\xi$ the mostly sulcus model (MSM).  
About 25\% of current sources in $R$ lie in gyri.  Therefore, 
if we set $\xi$=0.75, very few potentially true sources will be left out. 
It turns out that in the region where 1-$S_p$ is only slightly 
less than unity, setting $P$ to $R_\xi$ can offer the best result.

\begin{figure}[h]
\center
\includegraphics[height=1.8in]{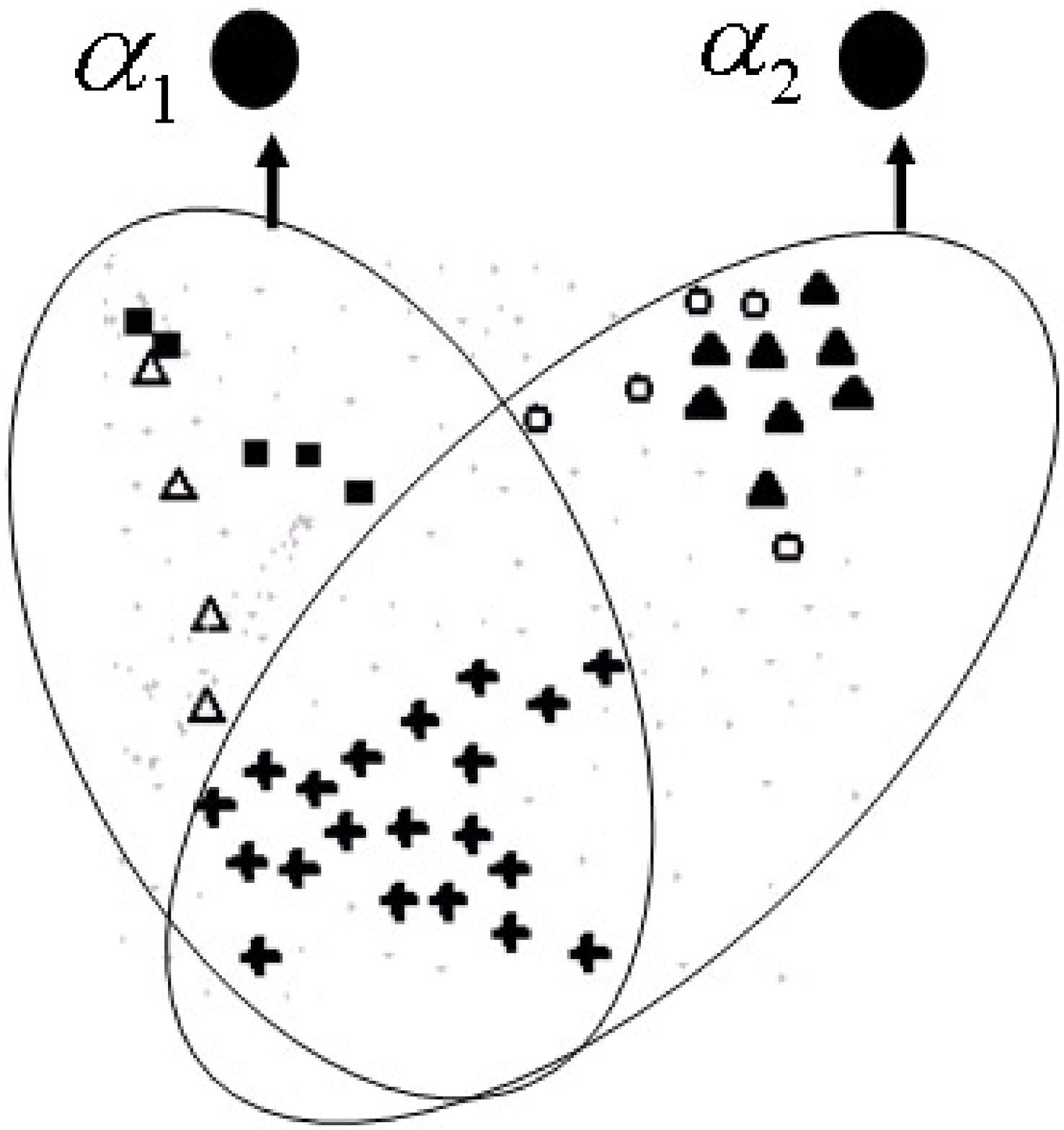}%{fig1bv3}
\caption{Illustration of backward filtering, when $M_S$ is composed of the two sensors,  
$\alpha_1$ and $\alpha_2$. Both $R_{\{\alpha_1\}}$  and 
$R_{\{\alpha_2\}}$ contain 3 clusters while $R_{SHM}$ contains only one cluster.}
  \label{f:SHM}
\vspace{-1cm}
\end{figure}

\sn \textbf{Backward Filtering}.   
Another way of reducing the pool of positives is to limit them to those clusters, with unit 
current strength,  whose 
$ \vert \mathcal{A}'\vert$ value is greater than a threshold value 
$A_0$ at  all the sensors in $M_S$.  This yields, for each $\alpha$, 
a reduced set with $N_\alpha < N_C$ 
clusters:  
\bgeq
R_>^{\{\alpha\}}=\{C_{u_\alpha}\vert u_\alpha=1,2,\cdots,N_\alpha\}_\alpha,\quad
\forall\ \alpha\in M_S. 
\label{e:reduced}
\edeq  
Now we let $R_{SHM}$, the pool of positives for the "simple head model" (SHM),  
be the intersection of the reduced sets,
\bgeq
R_{SHM}=\bigcap_{\alpha\in M} R_>^{\{\alpha\}}. 
\label{e:R_MEG}
\edeq
$R_{SHM}$ represents a 
coarse-grained, simplified cortex tailored to the MEG data at hand:
every source-cluster in $R_{SHM}$ has a relatively high probability of contributing  
significantly to all the sensors in $M_S$.  
A hypothetical case in which $M_S$ is composed of the two sensors,  
$\alpha_1$ and $\alpha_2$,  is depicted in \reffg{f:SHM}. 

%%%%%%%%%%%% {f:SHM} %%%%%%%%%%%%%%
%%%%%%%%%%%%%%%%%%%%%%%%%%%%%
For this paper $A_0$ is set to be 4.5 fT. Then the $N_\alpha$'s have values  
113, 134, 102, 126, 103, 119, and 64, respectively, for the 7 sensors in $M_S$, and the 
simple head model set $R_{SHM}$ contains 240 source currents, or about 11\% of the 
total number of current dipoles in the auditory region. 

\sn \textbf{The Maximum Entropy Method}. 
The maximum entropy (ME) method is a method for deriving the "best" solution in 
ill-posed problems  \cite{ME:ME,Clarke88,Jones89,Jones90,Csiszar91,Alavi93,
Khosla97,Besnerais99,He00,Gzyl02}. 
Generally, the equation that admits multiple solutions is treated as a 
constraint and, given a {\it prior} probability distribution of solutions, the method 
finds a {\it posterior} probability distribution by maximizing the relative entropy of the 
probability distributions.   When applied to the MEG case, 
\refeq{e:inverse_cluster} (or \refeq{e:inverse} without clustering) is used 
to constrain the posterior propbability distribution for a $\hat r$ that is the
``best'' ME solution, given {\bf m} (measured) and {\bf n} (presumed or otherwise 
obtained).  The procedure is tedious but standard and an implementation was  
reported in \cite{Pai2006}.   Here we only describe how the 
{\it prior} probability distribution of solutions are determined in this work. 
 %%%%%%%%%%% {f:Flowchart} %%%%%%%%%%%
\begin{figure}%[hbp]
\center
  \includegraphics[height=2.3in]{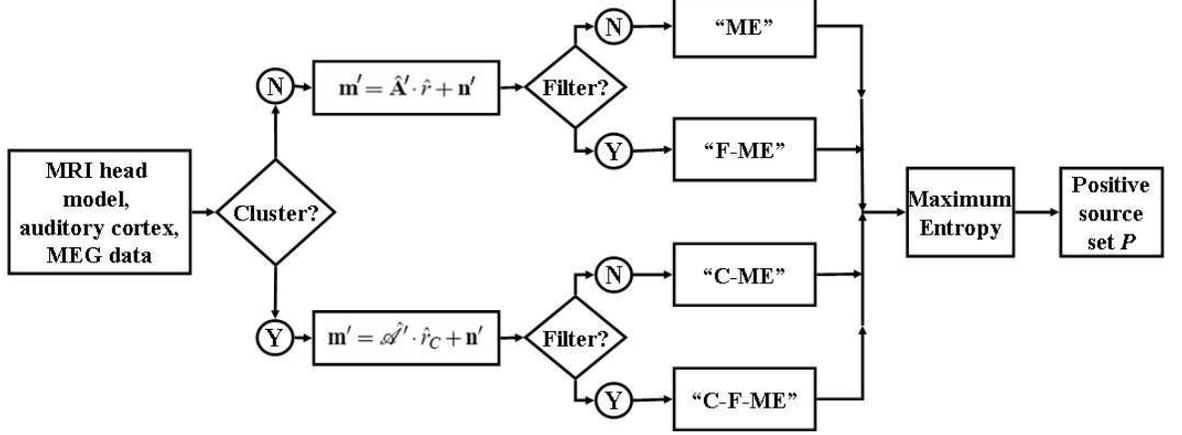}%{Flowchart}
  \caption{Data processing flowchart for producing a {\it prior} input for the maximum 
  entropy method.}
  \label{f:Flowchart}
\end{figure}
%%%%%%%%%%%%%%%%%%%%%% 

We test several procedures ranked by their levels of complexity: 
(1) Simple ME (ME). All the 2188 individual dipoles are 
included in the {\it prior} $P$. Here as in all other cases, 
in ME iteration involving sensors, only those in the "clear signal set" $M_C$ are included. 
(2) ME with clustering but not filtering (C-ME).   Cluster are treated as  
units of sources and all clusters are included in the {\it prior} $P$.
(3) ME with filtering but not clustering (F-ME).   Individual dipoles are 
treated as units of sources but only those in $R_\xi$ or $R_{SHM}$  
(whichever is the smaller set) are included in the {\it prior} $P$.
(4) ME with clustering and filtering (C-F-ME).  Cluster are treated as  
units of sources but only those in $R_\xi$ or $R_{SHM}$ are included in the {\it prior} $P$.
\reffg{f:Flowchart} is the flowchart for computation for the above procedures.  In each case the 
set of positives, $P$, hence $S_p$, is varied during the implementation of ME 
by a threshold on the strength of source dipoles for acceptance into the set.  
The exception is when the "F" procedure is taken.  In this case, when 0$<$$S_p$$\le$0.25, 
the set of positives, $P$, is directly set equal 
to $R_\xi$ as described in \refeq{e:R_MSM}, without going through the ME procedure.  
When $S_p$$>$0.25 the procedure is switched to SHM followed by ME.    
In addition, we compare ME and MNLS \cite{Method:MNLS}.  When MNLS is involved 
the flowchart is the same as given in \reffg{f:Flowchart}, with ME replaced by MNLS. 

\section{Results and discussions}

We report three preliminary tests exploring the properties of the procedures.  

\begin{figure}%[hbp]
\center
\includegraphics[height=2.0in]{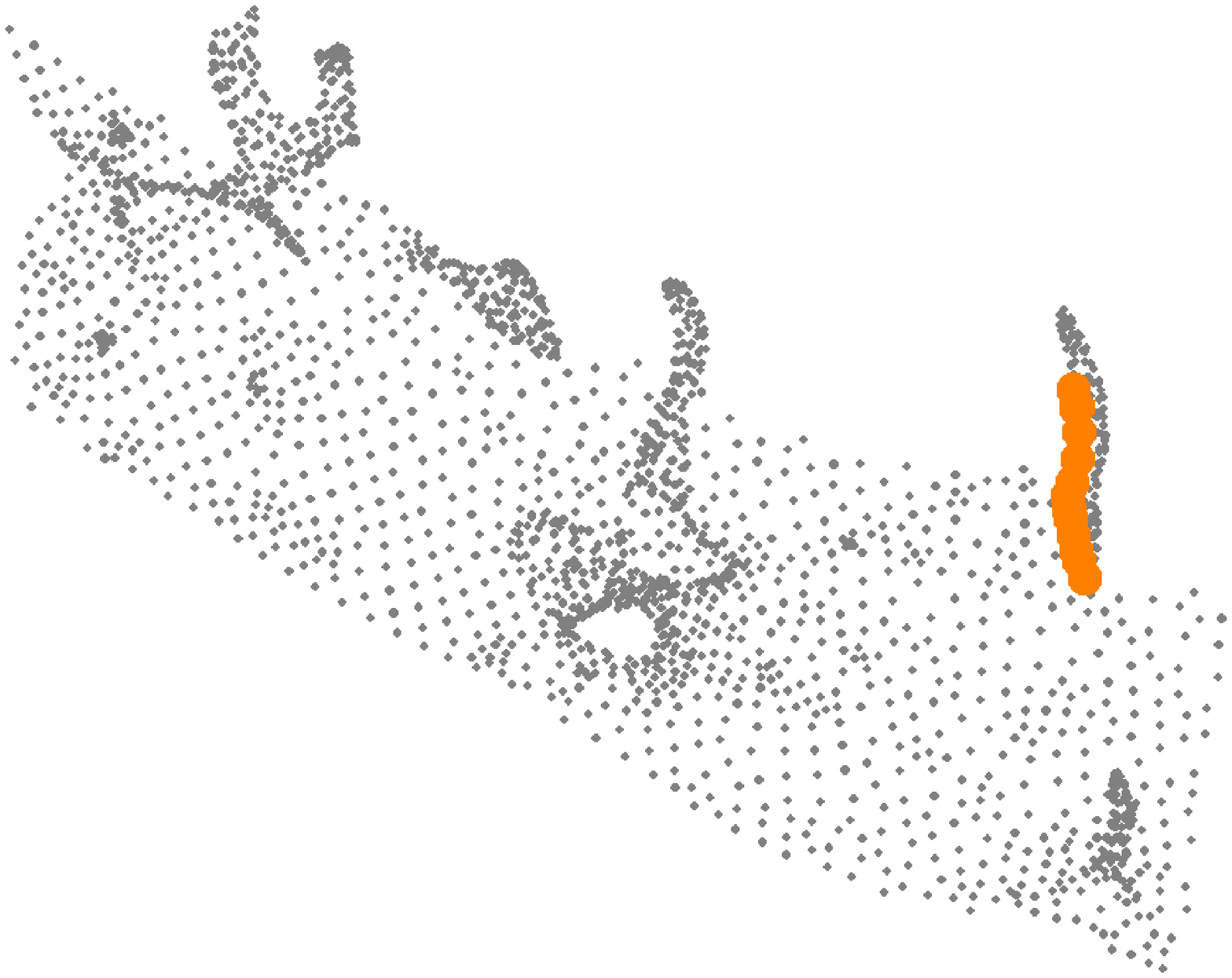}%{OneActualv1}\hspace{5pt} 
\includegraphics[height=2.2in]{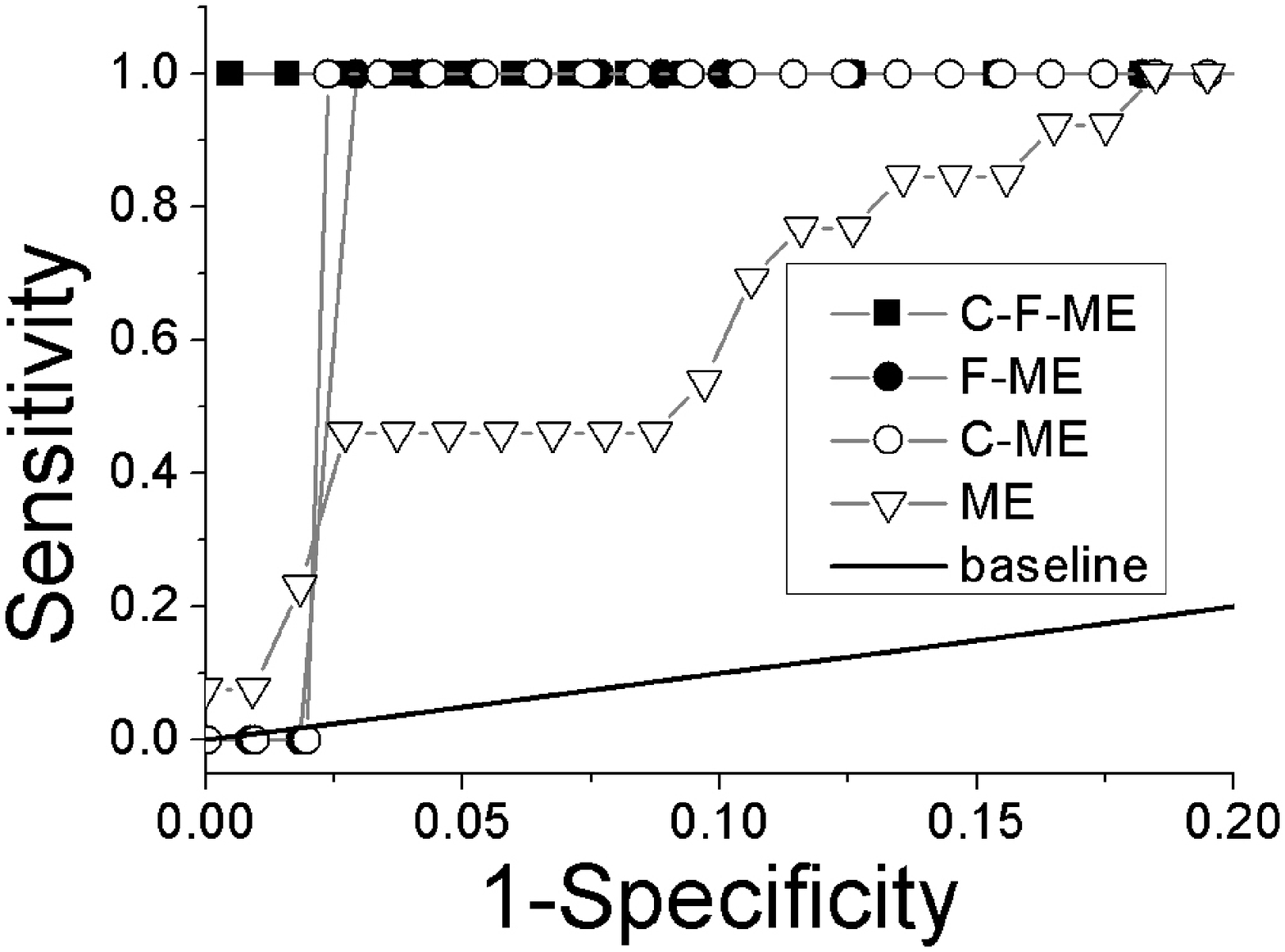}%{OneCluActROC}
  \caption{Left,  true cluster set of sources (solid circles)  in test 1.  
  Right, ROC plot for test 1.}
  \label{f:Test_one}
\end{figure}

\sn \textbf{Test 1}.  
The true set $T$ is one cluster containing 13 sources covering an 
area of 4 mm by 1 mm (\reffg{f:Test_one}L).  
The computed results, shown as an ROC plot, are given in \reffg{f:Test_one}R.   
%%%%%%%%%%% {f:Test_one} %%%%%%%%%%%
%%%%%%%%%%%%%%%%%%%%%%
Not shown are results 1-$S_p$$>$0.20, when all four procedures give $S_n$=1. 
It is seen that C-F-ME gives the best result: $S_n$=1 for the entire range of values 
for  1-$S_p$.  C-ME and F-ME are excellent when 1-$S_p$$\ge$0.025, but completely 
fail to identify the true sources when 1-$S_p$$<$0.025.  In comparison, ME performs not as well  
as C-ME and F-ME when 1-$S_p$$\ge$0.025 but is better when smaller values of 
1-$S_p$.  Note that the $R$ set contains 2188 sources and the $T$ set 13 sources.  So even 
at 1-$S_p$=0.025 there are still 54 false positive ($FP$) sources. 

\begin{figure}[hbp]
\center
\includegraphics[height=2.0in]{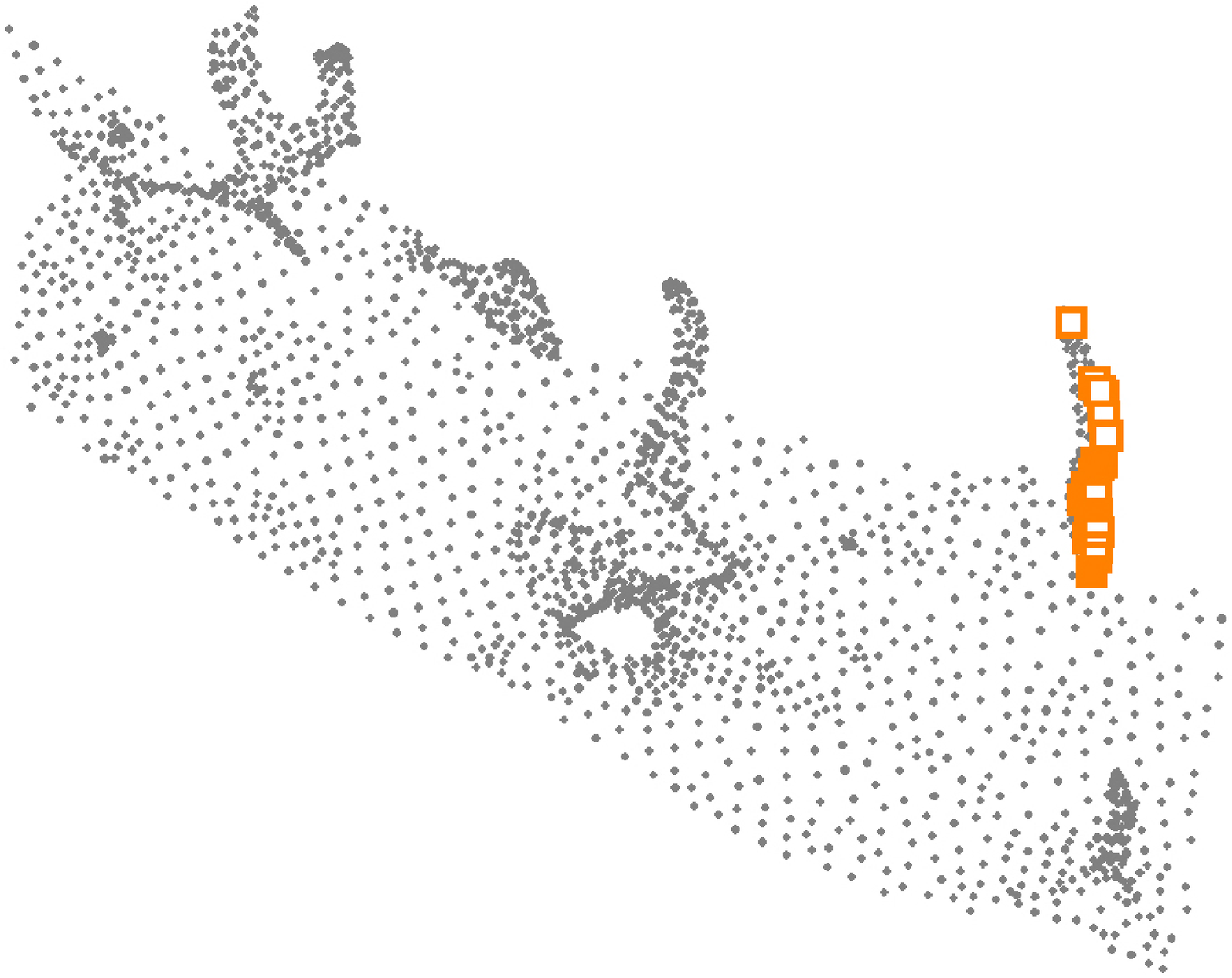}%{TwoActualv1}\hspace{5pt} 
\includegraphics[height=2.2in]{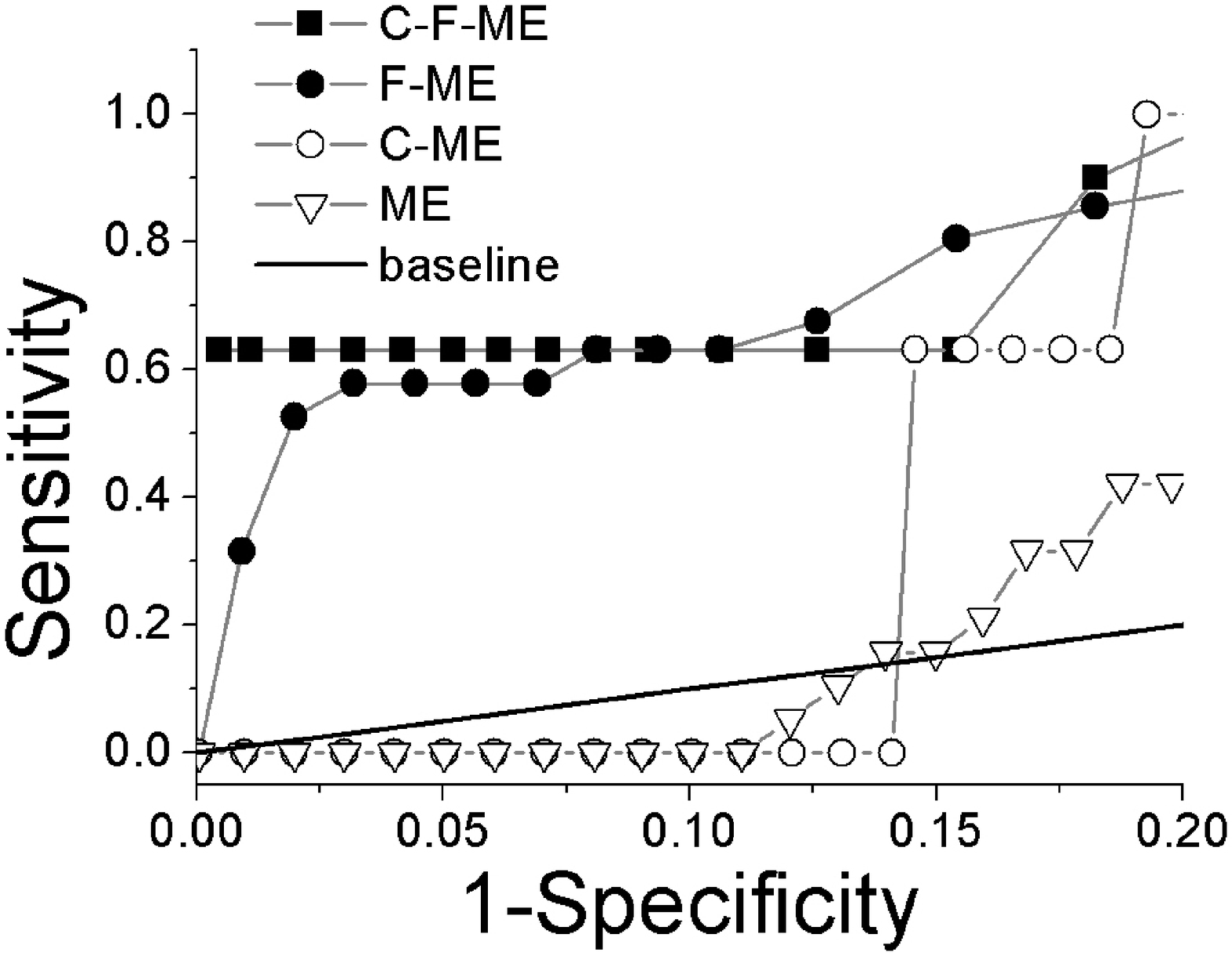}%{TwoCluActROC_1}
  \caption{Left, the two true cluster sets of sources (solid circles and hollow squares) in test 2. 
 Right, ROC plot for test 2. }
  \label{f:Test_two}
\end{figure}

\sn \textbf{Test 2}.  
The true set $T$ is composed of two clusters containing 12 and 7 sources, 
respectively, covering an area 5.5 mm by 1 mm (\reffg{f:Test_two}L).   
The ROC plot (\reffg{f:Test_two}R) shows 
better results are obtained when F is involved (F-ME and C-F-ME).  
When clustering is involved (C-ME and C-F-ME) changes in $S_n$ are discrete. 
This is because $P$ may take only take four values, 1 when $P$ contains both true 
clusters, 0.63 or 0.37 when it contains of the two, and 0 when it contains none. 
%%%%%%%%%%% {f:Test_two} %%%%%%%%%%%
%%%%%%%%%%%%%%%%%%%%%%
Recall that clustering simplifies 
the organization of the sources ($R$) but does not reduce the {\it prior} positives ($P$). 
F reduces the {\it prior} $P$ but does not simplify $R$.   C-F does both.  
That procedures with F are clearly better than those without highlights the 
paramount importance of a better {\it prior} in all but trivial situations when ME is employed.   

\begin{figure}%[hbp]
\center
  \includegraphics[height=2.0in]{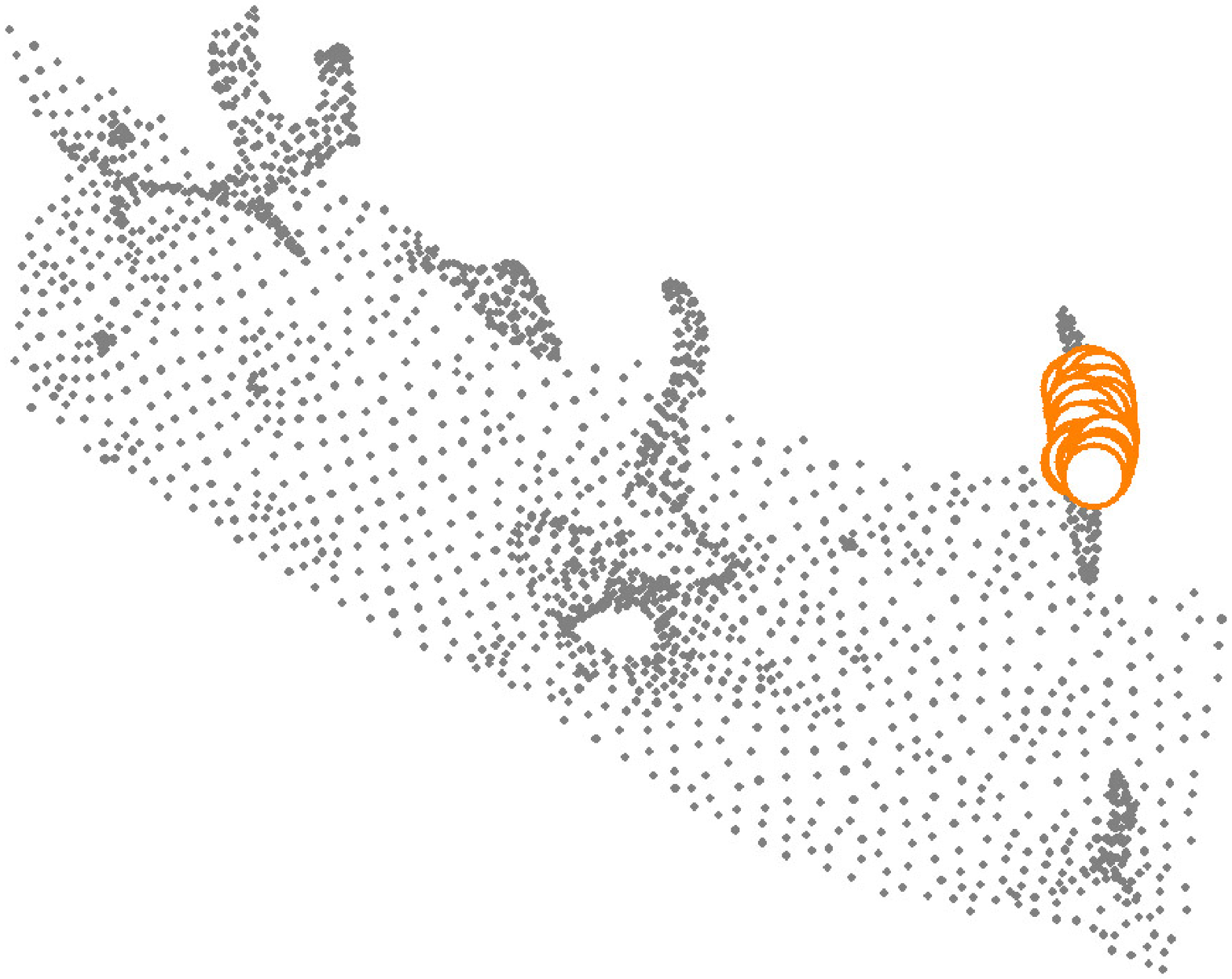}%{active_sources}\hspace{6pt} 
  \includegraphics[height=2.2in]{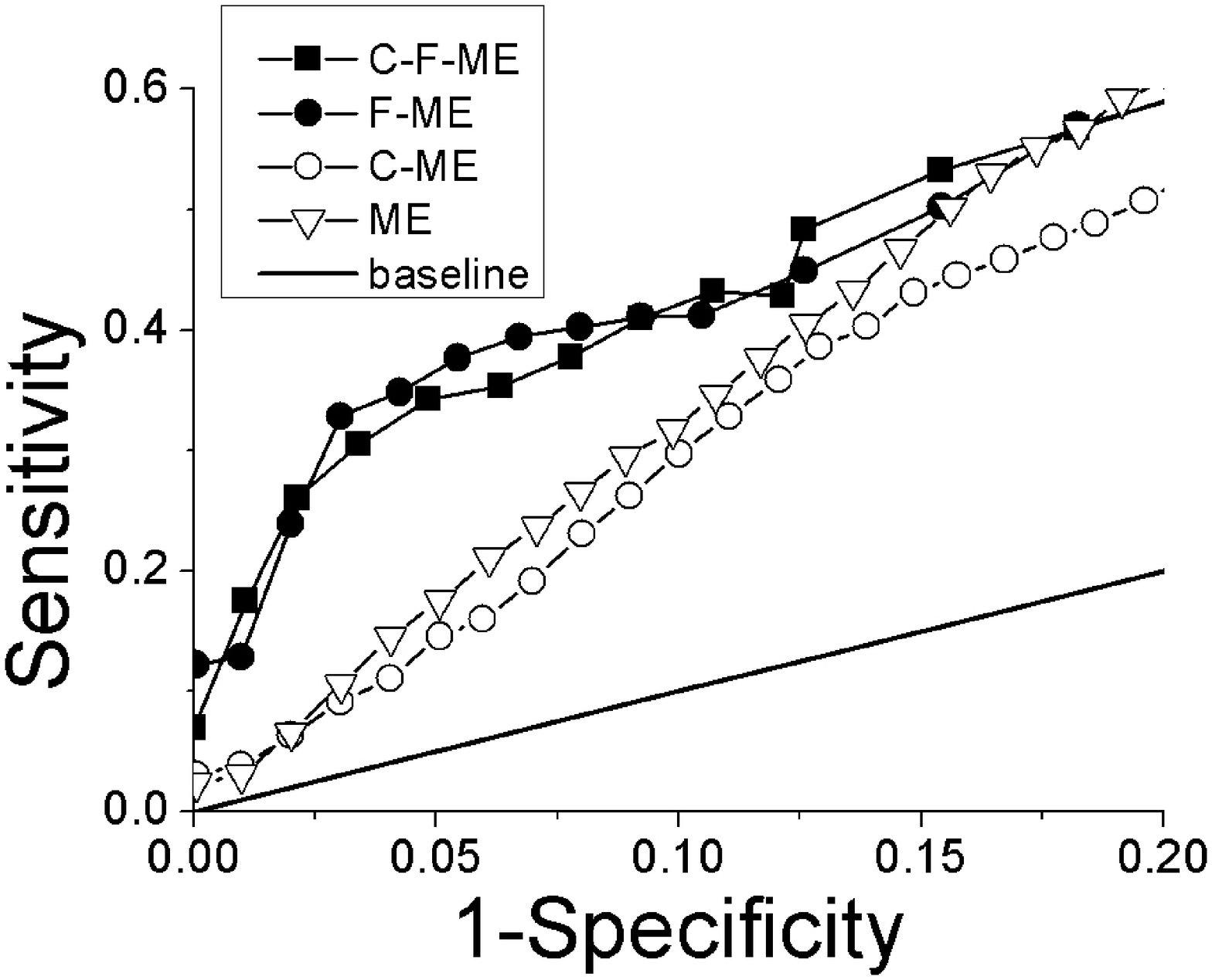}%{01ROC}
  \caption{Left, the 42 source dipoles in test 3. 
  Right, ROC plot for test 3; only results with 1-$S_p$$<$0.2 are shown.}
  \label{f:me4levelroc}
\end{figure}

\sn \textbf{Test 3}.  
The true set $T$ is composed of 42 active sources covering an area 
approximately 3 mm by 2 mm (\reffg{f:me4levelroc}L).  
%%%%----------figure---------------
%%%%----------figure---------------
They are distributed in eight clusters ($C_u$, $u$=1 to 8) containing  
18, 31, 15, 12, 9, 23, 18, and 17 sources, respectively, for a total of 143 sources.  
The intersection of these clusters with $T$ are 4, 11, 4, 3, 4, 4, 6, and 6 sources, respectively. 
If clustering is applied, the minimum value for 1-$S_p$ is 0.047 when $S_n$=1.   
The ROC shown in \reffg{f:me4levelroc}R shows that for this relatively complicated 
case simultaneous high $S_n$ and $S_p$ is difficult to achieve; we obtain 
$S_n$$<$0.6 for 1-$S_p$$<$0.20 in all procedures.  Here again, procedures with 
F, which generate better {\it priors}, yield more accurate positives than those without.  It is worth 
pointing out the accuracies of the positives given by F-ME and C-F-ME are very similar 
in almost the entire range of 1-$S_p$, but F-ME is decisively better than C-F-ME when 
1-$S_p$ is less than 0.01.  The last effect brings out an inherent weakness of clustering. 
If the active sources (the $T$ set) are spread out in more than one cluster and and if the 
union of the clusters is greater than the $T$ set, then a solution with a non-null $P$ 
($S_n$$\ne$0) and null $FP$ (1-$S_p$=0) is not possible.  
In contrast, such an outcome is at least possible without clustering. 
As it happened, for the case at hand, at 1-$S_p$$\approx$0, the $P$ set for C-F-ME 
is cluster $C_4$ with 12 sources, of which 3 belongs to $T$, yielding 
$S_n$=3/42=0.071 and 1-$S_p$=9/2146=0.0042.  

\begin{figure}[hbp]
\center
  \includegraphics[height=2.0in]{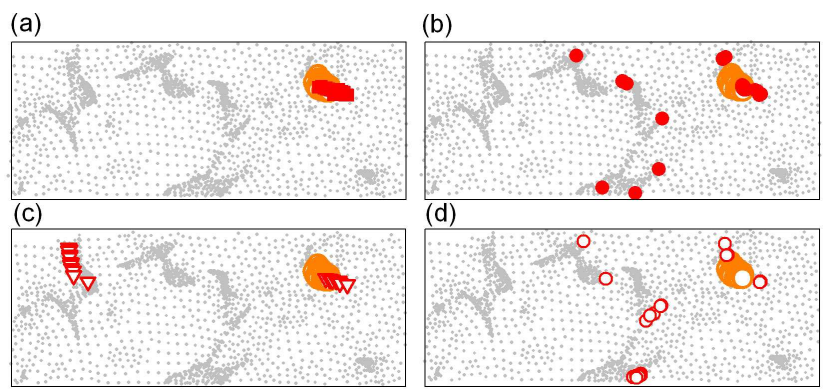}%{C_F_MEPositive}  %%%%
  \caption{The strong sets (21 strongest sources) in test 3 at 1-$S_p$$<$0.02.  
  In each case the gray (orange in color) blotch near the top-right corner is the $T$ set 
  of 42 active sources. 
(a) C-F-ME (solid squares).  (b) F-ME (bullets). 
(c) C-ME (triangles).  (d) ME (open circles).}
  \label{f:cluster_USHM_compared} 
\end{figure}

\reffg{f:cluster_USHM_compared} shows the  
21 strongest sources (the "strong set"), not necessarily all in the positive set, given 
%%%%%%%%%%% Plots %%%%%%%%%%%
%%%% clustering
 by the four procedures when 1-$S_p$$<$0.02 (the sets do not change much in this 
 range of 1-$S_p$).  In the case of C-F-ME (panel (a)), the strong set comes 
 from clusters $C_3$ and $C_4$, both of which lie in the vicinity of $T$.  
 When 1-$S_p$ is lowered (by raising the $P$-acceptance threshold) beyond a certain point 
 $C_3$ is eliminated, leaving only $C_4$ in $P$ in the situation discussed above.  
 In the case of F-ME (panel (b)), 
some individual sources in the strong set are not in the vicinity of $T$.  However, these 
are eliminated when 1-$S_p$ is lowered, and the remaining individual sources happen 
to have a $FP$ that is smaller than that in C-F-ME.  

\begin{figure}[hbp]
\center
  \includegraphics[height=2.4in]{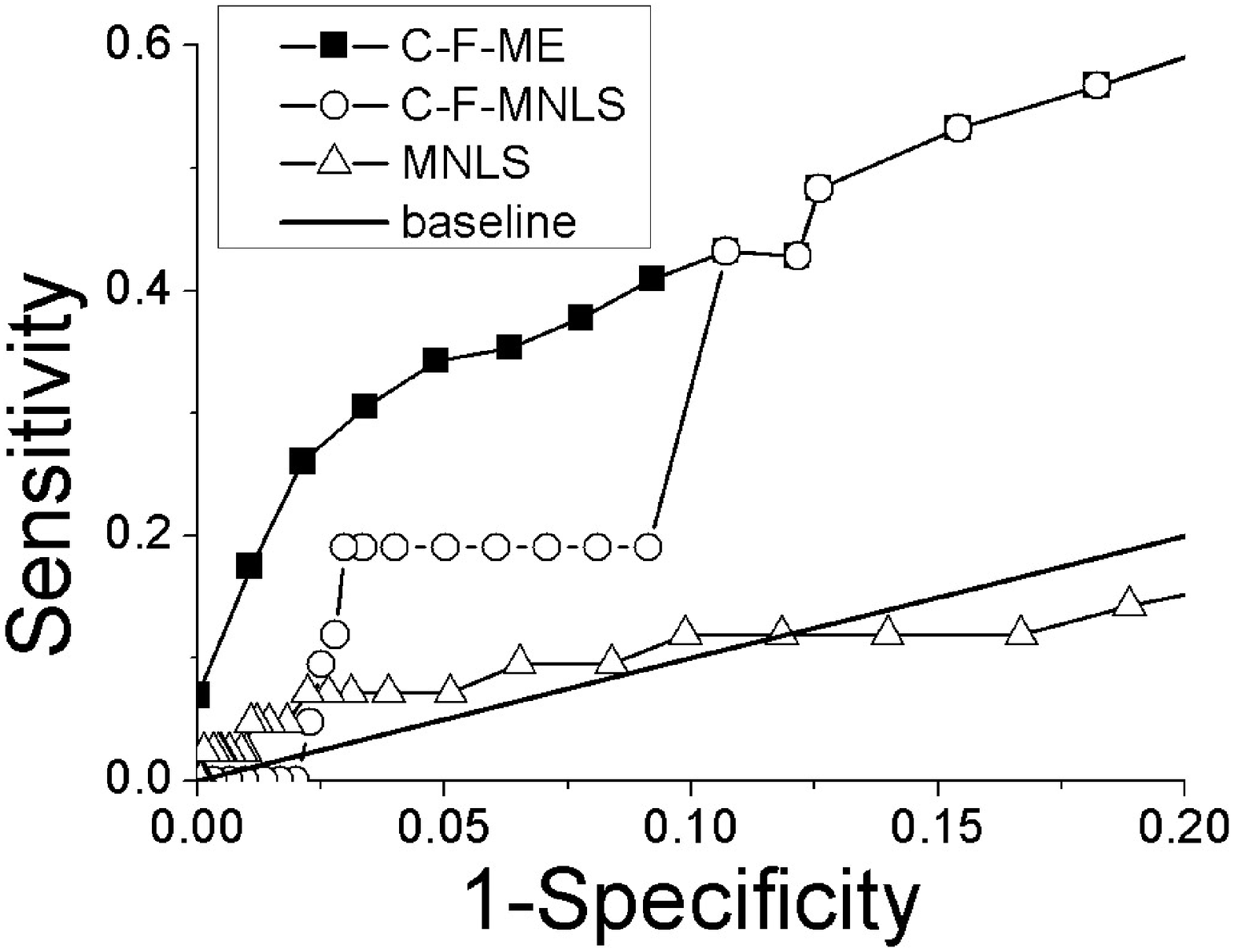}%{MEvsMNLS}\hspace{15pt}
  \includegraphics[height=2.4in]{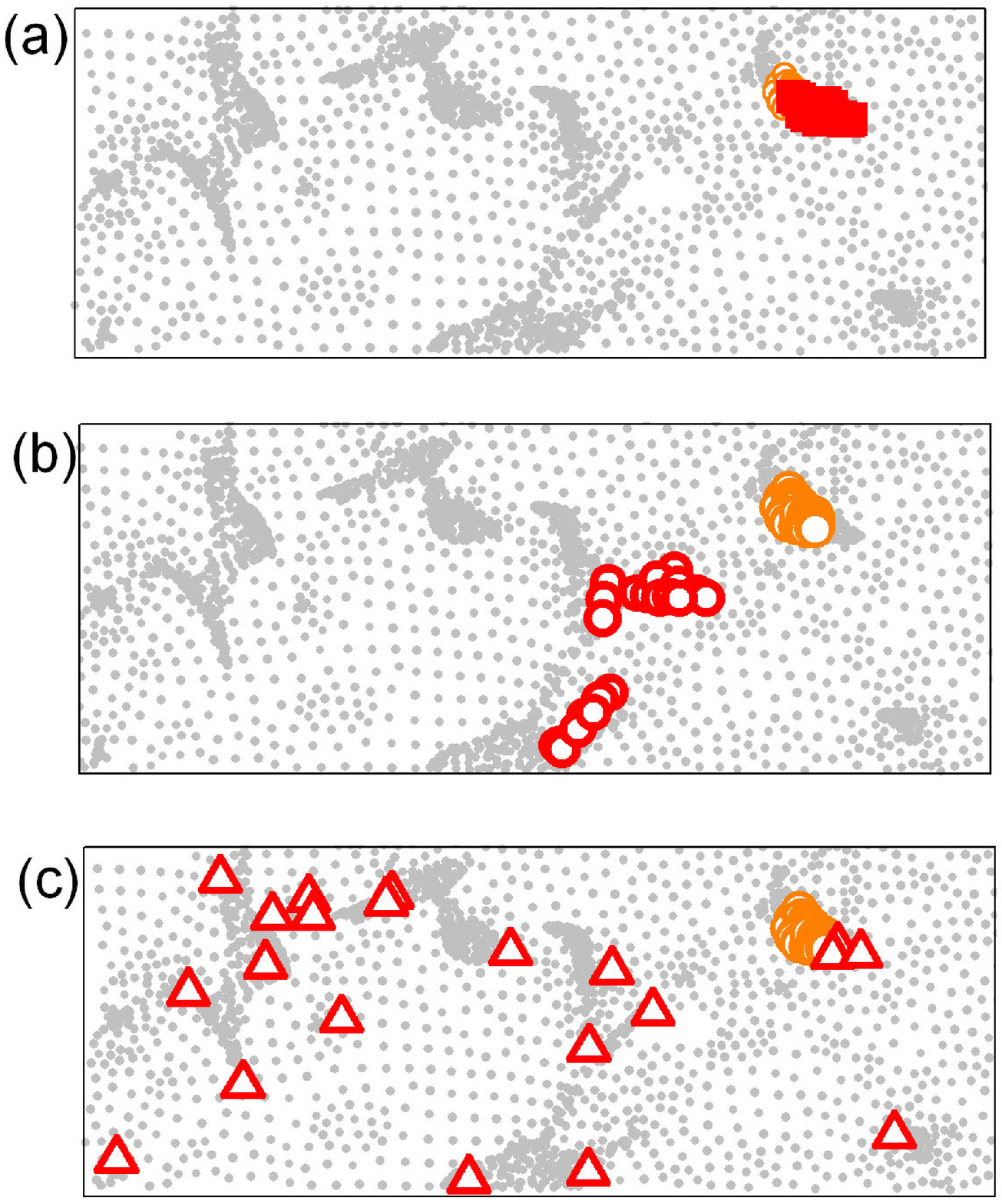}%{MEvsMNLSposi}
    \caption{Left, ROC plots for MNLS, C-F-MNLS, and C-F-ME.  Set $T$ is 
    the 42 active sources of test 3. Right, the strong set (dark symbols) at 1-$S_p$$<$0.02. 
    (a) C-F-ME (solid square); (b) C-F-MNLS (open circle);  (c) CMNLS (triangle). 
  $T$ is the set of grey (orange) circles near the top-right corner in panels (a-c).}
  \label{f:mevsmnlsposi}
  %\vspace{-0.5cm}
\end{figure}

\sn \textbf{ME versus MNLS}. 
We compare the effectiveness of the MNLS procedure against ME using the true set of test 3.  
\reffg{f:mevsmnlsposi}L shows the ROC plots for C-F-ME 
(same plot as in \reffg{f:me4levelroc}), C-F-MNLS and MNLS.  
We observe that MNLS is worse than ME (\reffg{f:me4levelroc}), 
C-F-MNLS is better than MNLS, and C-F-ME is better than C-F-MNLS. This shows, 
at least for the case tested, clustering is also beneficial to MNLS and, other things 
being equal, ME is more effective than MNLS.  In the last instance the $FP$'s of MNLS cover 
a large area of the auditory cortex.  
%%%%%%%%%% {f:mevsmnlsposi} %%%%%%%%%%%%%
%%%%%%%%%% {t:MEvsMNLS} %%%%%%%%%%%%%
The strong sets in the three procedures are given in \reffg{f:mevsmnlsposi}R.  
If we take these dipoles 
as the positive set then we obtain the results given in Table \ref{t:MEvsMNLS}.
%%%%%%%%%%%%%  
\begin{table}[hbt]
\center
\begin{tabular}{lrcccc}
\hline
\hspace{1.2in}&\hspace{1.2in}&\hspace{0.5in} &\hspace{3pt} C(luster) \hspace{3pt}&
\hspace{5pt} F\hspace{5pt} &\hspace{5pt} C-F\hspace{5pt} \\
\hline
$S_{n}$ & ME & 0.000 & 0.048 &0.119 & 0.167    \\
   & MNLS  & 0.000   & 0.000  &0.048 & 0.000 \\
   \\       
$1-S_{p}$ &ME  & 0.0096  & 0.0087 &0.0073 & 0.0064 \\
   &MNLS    & 0.0096   & 0.0096  &0.0087 & 0.0096\\
\hline
\end{tabular}
\caption{Comparisons of ME and MNLS using the set of 42 active sources of test 3.  
In each procedure the positive set is composed of the 
21 current dipoles with the strongest currents. The third column gives results with neither 
clustering nor F.}
\label{t:MEvsMNLS}
\end{table}
%%%%%%%%%%%%%%%%%

In summary, we have employed several simple cases to illustrate  
that even when we may use fMRI data to tell us where the general 
region of the source currents are, the nature of the inverse problem is such that 
the challenge to precisely pinpoint sources located in a small area is still great. 
We showed that clustering tends to limit the area covered by false positives, 
and filtering is effective for generating better {\it priors} for ME.  
When we use ME implemented by these 
procedures, we can achieve part partial success in pinpointing sources concentrated in an 
area the size of a few mm across, comparable to the spatial resolution of fMRI \cite{Moonen99}, 
while limiting the spatial distribution and 
number false positives.  Considering that the area of the 
active sources is miniscule compared to the auditory cortex, which is itself yet much 
smaller than cortical surface covered by the sensors receiving  
clear or even strong signals, the achieved level of success reported here, even as it still 
leaves much to be desired, is a vast improvement 
over what can be inferred simply from the pattern of sensors with strong signals. 

This work is supported in part by grant nos. 95-2311-B-008-001 
and 95-2911-I-008-004 from the National Science Council (ROC).

%%%%%%%%%%%%%%%%%%%%%%%%%%%%%%%%%%%%%%%%%%%%%%%%
%% BACKMATTER
%%%%%%%%%%%%%%%%%%%%%%%%%%%%%%%%%%%%%%%%%%%%%%%%

\end{document}